\documentclass[journal=jacsat,manuscript=article]{achemso}

\usepackage[version=3]{mhchem} % Formula subscripts using \ce{}

\author{Milena De Giorgi}
\affiliation{CNR NANOTEC-Institute of Nanotechnology, Via Monteroni, 73100 Lecce, Italy}

\author{Mohammad Ramezani}
\affiliation{Dutch Institute for Fundamental Energy Research, DIFFER, P.O. Box 6336, 5600 HH Eindhoven, The Netherlands}
\email{m.ramezani@differ.nl}

\author{Francesco Todisco}
\affiliation{CNR NANOTEC-Institute of Nanotechnology, Via Monteroni, 73100 Lecce, Italy}

\author{Alexei Halpin}
\affiliation{Dutch Institute for Fundamental Energy Research, DIFFER, P.O. Box 6336, 5600 HH Eindhoven, The Netherlands}

\author{Davide Caputo}
\affiliation{CNR NANOTEC-Institute of Nanotechnology, Via Monteroni, 73100 Lecce, Italy}
\alsoaffiliation
{Dipartimento di Fisica, Universit\'{a} del Salento, Strada Provinciale Lecce-Monteroni,  Campus Ecotekne, Lecce 73100, Italy}

\author{Antonio Fieramosca}
\affiliation{CNR NANOTEC-Institute of Nanotechnology, Via Monteroni, 73100 Lecce, Italy}
\alsoaffiliation
{Dipartimento di Fisica, Universit\'{a} del Salento, Strada Provinciale Lecce-Monteroni,  Campus Ecotekne, Lecce 73100, Italy}

\author{Jaime Gomez Rivas}
\affiliation{Dutch Institute for Fundamental Energy Research, DIFFER, P.O. Box 6336, 5600 HH Eindhoven, The Netherlands}
\alsoaffiliation{Department of Applied Physics and Institute for Photonic Integration, P.O. Box 513, 5600 MB, Eindhoven, The Netherlands}

\author{Daniele Sanvitto}
\affiliation{CNR NANOTEC-Institute of Nanotechnology, Via Monteroni, 73100 Lecce, Italy}
\alsoaffiliation{INFN Sezione di Lecce, 73100 Lecce, Italy}
\email{daniele.sanvitto@nanotec.cnr.it}

\title{Interaction and coherence of a plasmon-exciton polariton condensate}

\abbreviations{IR,NMR,UV}
\keywords{American Chemical Society, \LaTeX}

\begin{document}


%\begin{tocentry}
%\includegraphics[width=3.5in]{TOC2.pdf}
%\end{tocentry}

\begin{abstract}
Polaritons are quasiparticles arising from the strong coupling of electromagnetic waves in cavities and dipolar oscillations in a material medium. In this framework, localized surface plasmon in metallic nanoparticles defining optical nanocavities have attracted increasing interests in the last decade. This interest results from their sub-diffraction mode volume, which offers access to extremely high photonic densities by exploiting strong scattering cross-sections. However, high absorption losses in metals have hindered the observation of collective coherent phenomena, such as condensation. In this work we demonstrate the formation of a non-equilibrium room temperature plasmon-exciton-polariton condensate with a long range spatial coherence, extending a hundred of microns, well over the excitation area, by coupling Frenkel excitons in organic molecules to a multipolar mode in a lattice of plasmonic nanoparticles. Time-resolved experiments evidence the picosecond dynamics of the condensate and a sizeable blueshift, thus measuring for the first time the effect of polariton interactions in plasmonic cavities. Our results pave the way to the observation of room temperature superfluidity and novel nonlinear phenomena in plasmonic systems, challenging the common belief that absorption losses in metals prevent the realization of macroscopic quantum states.
\end{abstract}

\noindent{keywords: polaritons, plasmonics, plexcitons, Bose-Eistein condensation}\\

The study of  exciton-polaritons, \textit{i.e.}, a bosonic quasiparticle formed by the strong coupling between photons and excitons,  has attracted a great attention in recent years on both experimental and theoretical sides.  The research in this field has mainly focused on inorganic semiconductors, showing fascinating phenomena such as condensation,\cite{Kasprzak:2006jy} superfluidity,\cite{Amo:2009,Amo:2009bl} quantized vortices,\cite{Lagoudakis:2008} and nonlinear dynamics.\cite{Savvidis:2000}  
Observation of these phenomena has been obtained very recently also at room temperature in organic-based microcavities\cite{KenaCohen:2010cy, Plumhof:2013bn, Daskalakis:2014ex, Lerario:2017}.  These have been some of the major breakthroughs in this research field because they permit to overcome cryogenic temperature limitations to the application of exciton-polaritons in optical devices.

In this context, a promising route for the exploitation of exciton-polariton physics at the nanoscale is offered by Localized Surface Plasmons (LSPs) characterized by unique properties such as the confinement of light in mode volumes far below the diffraction limit ($V\ll\lambda^3$) and the strong electromagnetic (EM) field enhancement.  These systems have already shown many interesting features even in the strong coupling regime, where plasmon-exciton-polariton (PEP, or some times called plexciton) quasiparticles are formed\cite{Hakala:2009, Chikkaraddy:2016,Zengin:2013,Torma:2015,Todisco:2017}. 
Among these systems, particular attention is devoted to the study of two-dimensional (2D) lattices of metallic nanoparticles which support collective plasmonic modes, known as Surface Lattice Resonances (SLRs).\cite{Wang:2017} SLRs arise from the coherent radiative coupling of LSPs of individual nanoparticles with diffractive modes propagating in the plane of the array, the so-called Rayleigh anomalies,\cite{zou:2004,Kravets:2008,Rodriguez:2011} and are characterized by a strong suppression of losses (higher quality factor) with respect to individual nanoparticle LSPs, at the expense of a less confined electromagnetic field (larger mode volume)~\cite{zou:2004, Vecchi:2009prl, Humphrey:2014}. As hybrid modes arising from a coherent coupling of LSPs, SLRs can maintain the strong EM field enhancement typical of plasmonic nanoparticles\cite{Todisco:2016,Ramezani:2016}, while simultaneously extending over the whole lattice area. 
PEPs in 2D lattices have shown peculiar features due to the unique dispersion of SLRs \cite{Rodriguez:13, Rodriguez:2013prl, Vakevainen:2014, Todisco:2015}, which make them similar to exciton polaritons in semiconductor microcavities. For these reasons, the possibility to achieve PEP condensation in these systems has been predicted.~\cite{Rodriguez:2013prl}

Although polariton condensates and photon lasing have many properties in common such as excitation density threshold, coherence of the emitted light and spontaneous polarization, they are intrinsically very different\cite{Snoke:2012, Yamamoto:2017}: polariton condensation arises from the Bose stimulation of quasiparticles into the ground state,  whereas photon lasing is due to the stimulated emission of photons into the resonator mode. As a result, the threshold for a polariton condensate can be even orders of magnitude lower than that one of a photon laser. Moreover, condensates preserve a uniform single spatial mode even with a large pump spot size and,  more importantly,  give to photons a missing ingredient, which is fundamental in the realization of all optical and quantum computation devices: the interaction. As an example, phenomena like superfluidity\cite{Amo:2009,Amo:2009bl,Lerario:2017} can only be observed in presence of interactions while they are absent in standard lasing.

Very recently a condensate of photons based on SLR modes in the weak coupling regime has been claimed showing hint of a thermalisation due to exchange of energy between the SLR modes and the molecular dipoles\cite{Torma:2018}. On the other hand, in strong coupling regime,  thermalization and cooling of polaritons has been observed in an array of silver nanoparticles covered with organic molecules. This has been shown to lead to a saturation of the strong coupling at large pumping fluences, and represented the first demonstration of suitability of PEPs in a plasmonic based system towards quantum condensation, even if no evidences of boson stimulation were reported\cite{Rodriguez:2013prl}. More recently, a similar platform has shown PEP lasing\cite{Ramezani:2017}, although the absence of evidences for the formation of a quasi-long-range spatial coherence and interactions has limited the discrimination between lasing and condensation.

In this work, we demonstrate that by strongly coupling molecular excitons to a SLR in a 2D array of silver nanorods (NRs), where the exciton interactions play a significant role, an out-of-equilibrium PEP condensate with a large spatial coherence  is observed. This is proved by the observation of exciton polariton interactions resulting in a blueshift of the condensate energy as large as 2.5 meV. A further confirmation comes from the spatial coherence length in the plasmonic lattice that is shown to extend to much longer distances than the pump spot size. These results shed new light on plasmonics, demonstrating the feasibility in achieving room temperature condensation of exciton-polaritons in open plasmonic cavities and paving the way for new active plasmonic devices based on hybrid interacting quantum fluids.

\section*{Non-equilibrium condensation}

The optical mode in our system is provided by a 2D array of silver nanorods (NRs) where the LSPs of the individual nanoparticles are coherently coupled with each other through the diffraction orders propagating in the plane of the array, thus sustaining SLRs. The NRs array, whose schematic representation and scanning electron micrograph are shown in Figure 1a, has been fabricated using conformal imprint lithography onto a glass substrate (see Methods section). The long and short pitches of the lattice are 380 nm and 200 nm, respectively. The NRs are 200 nm long, 70 nm wide, and 20 nm high. We refer to the y-axis as the direction parallel to the long axis of the NRs, as indicated in Figure 1a. A 260 nm thick layer of PMMA doped with a Rylene dye was spin coated on top of the array. The absorption spectrum of this dye (shown in the Supporting Information), has one main electronic transition at $E_{X_{1}}$=2.24 eV and a vibronic replica at $E_{X_{2}}$=2.41 eV as indicated in Figure 1b (black dashed lines) where angle-resolved extinction measurements of the sample are shown as a function of the incident photon energy and the in-plane wave vector parallel to the short NRs axis ($k_{x}$), with incident light polarized along the long axis of the nanoparticles. The dispersion of the bare SLR in the absence of the molecules is indicated by blue dashed line, as resulting from the coupling of the (0,$\pm$1) lattice mode with the LSP of the NRs.  The LSP resonance of the particles is visible in Fig. 1(b) as a flat band with E = 2.36 eV. On the other hand the energies associated to the electronic transition and the vibronic replica of the molecules are shown with horizontal black dashed lines. The lower polariton (LP) band induced by the strong coupling of these three resonances can be calculated by diagonalizing the three-level Hamiltonian, which includes the SLR and the coupling to the electronic transition and first vibronic replica of the molecules (see Supporting Information). The first two modes obtained by this diagonalization have been plotted in Figure 1b and correspond to the lower PEP band and middle PEP band (solid green lines).  The agreement between the model and the extinction measurement is good for lower-polariton. However, for the middle polariton due to the presence of multiple modes at this energy range the quality of the fit is poor. The associated Rabi splitting is $\Omega_R \simeq 200$  meV. Note that at slightly higher energies (E = 2.12 eV), another dispersive band appears due to the presence of a guided mode in the molecular layer.
A peculiar property associated to the LP band is that under symmetric illumination ($k_{x}$= 0 $\mu m^{-1}$) the sample becomes nearly transparent, i.e., the transmission increases and the extinction approaches zero. The origin of this behavior has been widely investigated and explained by the multipolar nature of LSP modes coupled with the lattice,~\cite{Giannini:2010} whose electromagnetic field pattern does not couple efficiently to far-field radiation. This results in a lower extinction and much longer photon lifetimes.\cite{Lienau:2005prl,SRK:2011prx, Abass:2014acs} The corresponding electromagnetic near-field distribution is shown in Figure 1c as evaluated from finite difference time domain (FDTD) simulations in a unit cell of the lattice at E = 2.056 eV and $k_{x}$ = 0 $\mu m^{-1}$. Here, the black arrows correspond to the real part of the electric field along x- and y-direction, evaluated in a plane positioned at half height of the NR. A coupled multipolar field distribution is apparent from the field mapping, thus confirming the observed reduced polarizability of the array for the incident wave with this wave vector and frequency.

\begin{figure}
\begin{center}
\includegraphics[width=6.5in]{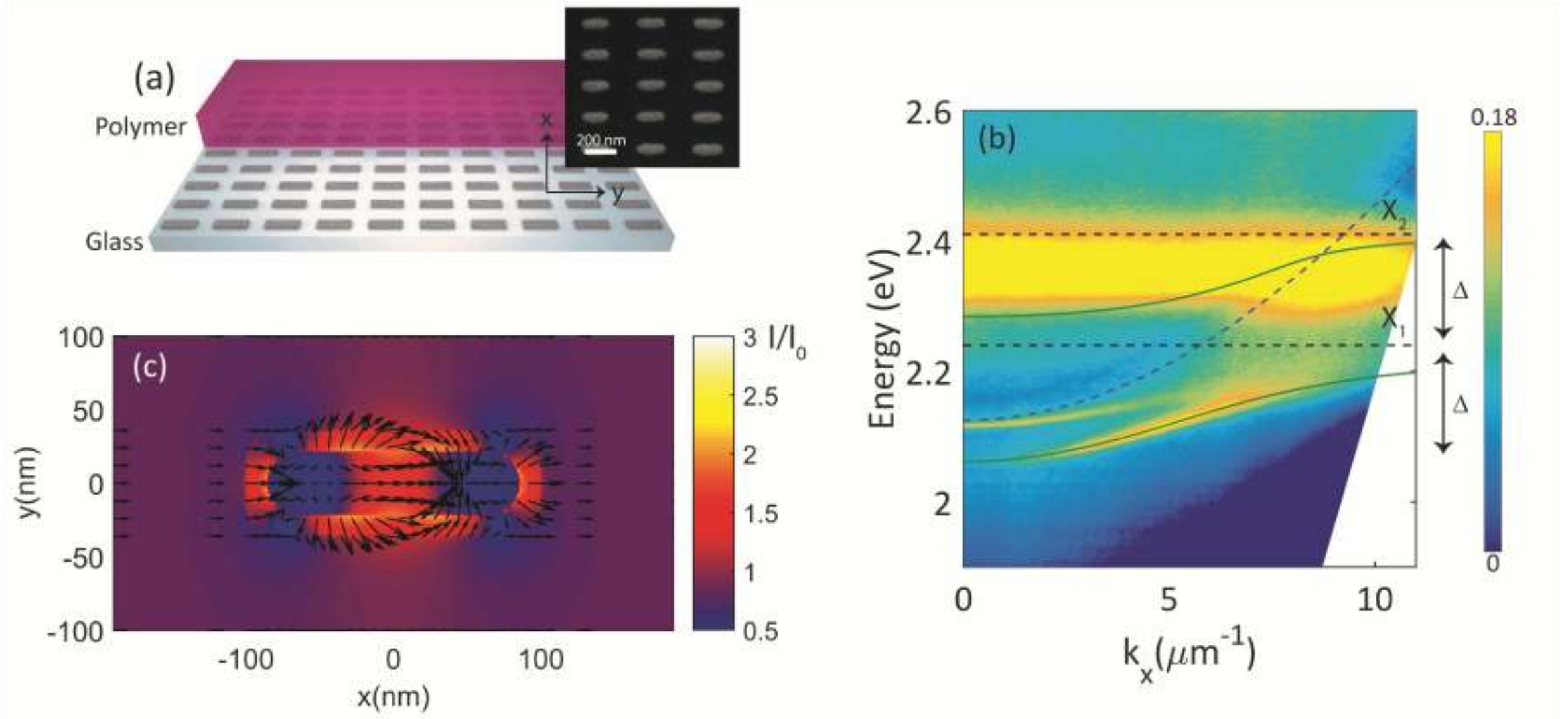}
\end{center} 
\caption{\textbf{PEP formation in a plasmonic lattice.} (a) Schematic representation and SEM image of the array of silver NRs. We indicate as y-axis the direction parallel to the long axis of the NRs. (b) Measured sample extinction and calculated dispersion, as obtained by solving a three-level Hamiltonian, showed as a function of the in-plane wavevector component parallel to the nanorods short axis, k$_{x}$. The black dashed lines indicate the energy of the electronic transition ($X_{1}$) and first vibronic replica ($X_{2}$) of the dye molecules. The blue dashed curve corresponds to the dispersion of the SLR. Green solid curves are the dispersions of the lower and middle PEPs. (c) Simulation of the total electric field intensity in the xy plane of a unit cell of the array illuminated by a plane wave with E = 2.056 eV and at $k_{x}=$ 0 $\mu m^{-1}$. The black arrows correspond to the real part of the electric field vector along x- and y-direction.}
\label{fig:Fig1} 
\end{figure}

To characterize the photoluminescence (PL) properties, we excited the sample using a non-resonant pulsed laser (100 fs pulse duration, $E_{exc}=2.48$ eV and 1 kHz repetition rate) at normal incidence, polarized along the y-direction. The dependency of the photoluminescence peak intensity as a function of absorbed pump fluence and the change in the emission linewidth are shown in Fig. 2a,b. A clear transition to the nonlinear regime is observed above the threshold power of $P_{th}$ =20 $\mu$J/cm$^2$,  resulting also in an enhanced temporal coherence due to the reduction of the linewidth. The PL dispersion as a function of $k_{x}$ is displayed in Figures 2c,d for pump fluences below ($P= 0.8P_{th}$) and above ($P= 1.1P_{th}$) threshold, respectively. At pump fluences below threshold (Fig. 2c), the PL dispersion follows the dispersion of the LP band shown in Fig. 1b. As the pump fluence is raised above $P_{th}$, the PL dispersion collapses into a sharp peak centered at E= 2.04 eV and $k_{x}=0$ $\mu$m$^{-1}$, as displayed in Fig. 2d ($P=1.1P_{th}$). One of the peculiar properties of organic-based polariton lasing is the presence of the vibrational progressions on the individual molecules as an efficient relaxation channel for exciton-polaritons.~\cite{KenaCohen:2010cy,mazza:2009prb,Ramezani:2017,Mazza2013} The presence of this relaxation channel helps PEPs to condense more efficiently at an energy set by the vibronic quanta of the molecules, $E_{X_{i}}$ showed in Fig.1 ($\Delta$= $E_{X_{2}}$-$E_{X_{1}}$=170 meV in our case).

The real space emission pattern also changes radically below and above threshold, as shown in Fig. 2e,f. While the PL is homogeneously distributed over the excited area below threshold (Fig. 2e), a structured stripe-like pattern extended along the y-direction arises at $P>P_{th}$ (Fig. 2f). The random allocation of the stripes across the emission pattern can be explained by sample imperfections and inhomogeneities as can be inferred by the different emission patterns (not shown) obtained on different regions of the sample.

\begin{figure}
\begin{center}
\includegraphics[width=5in]{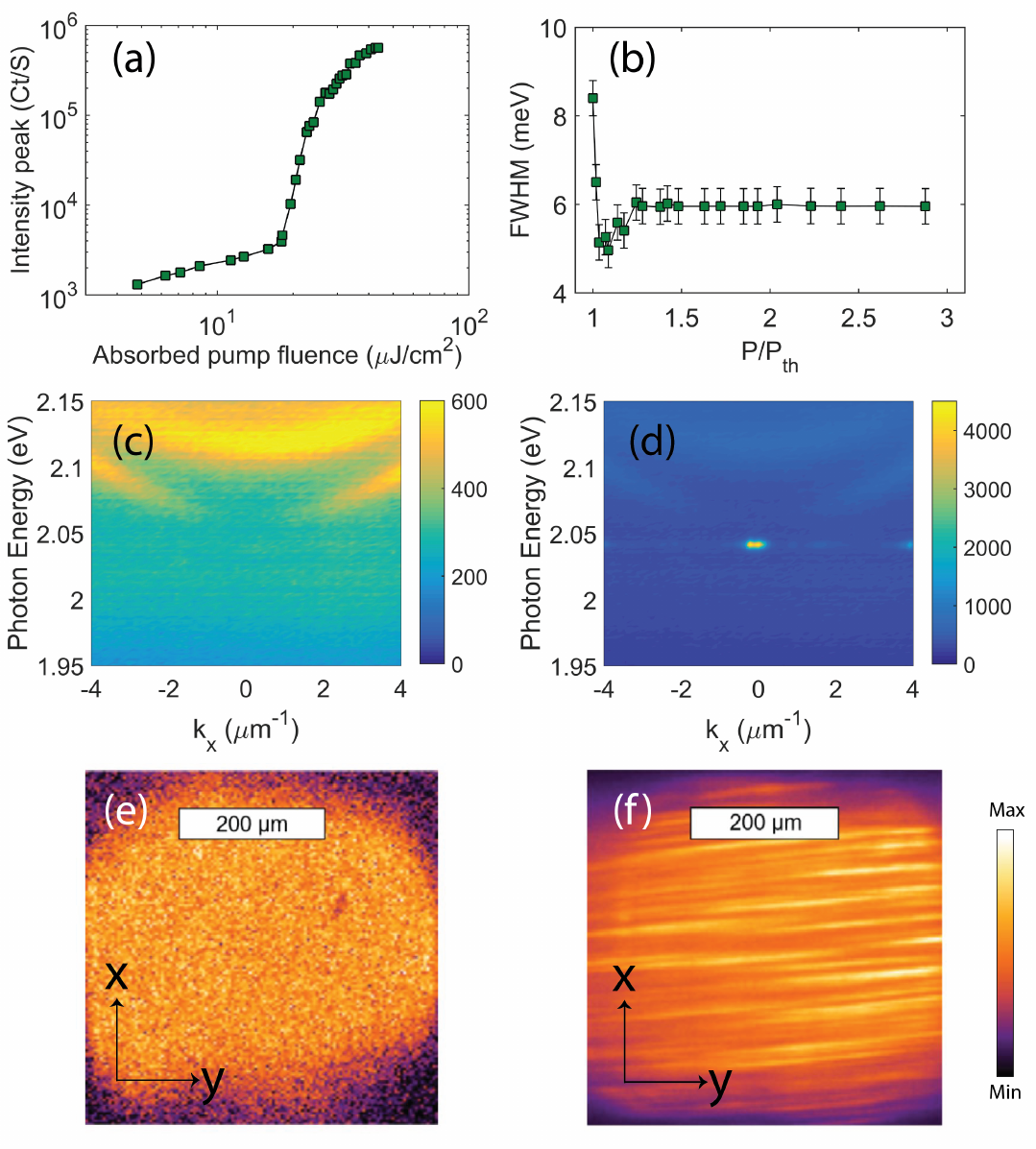}
\end{center} 
\caption{\textbf{Photoluminescence properties of PEPs below and above threshold.}  (a) Photoluminescence peak intensity versus the absorbed pump fluence. (b) The linewidth of the emission peak crossing the threshold. (c-d) Normalized angle-resolved photoluminescence for pump fluences (c) below ($P= 0.8P_{th}$) and (d) above ($P=1.1 P_{th}$) the condensation threshold. The false-color maps represent the emitted intensity in a linear scale. (e-f) Real-space emission maps showing homogeneous and structured emission, below (e) and above (f) threshold, respectively.}
\label{fig:Fig2} 
\end{figure}

\section*{Time-resolved photoluminescence}

To gather fundamental insights into the nature of this coherent emission, we performed time- and energy-resolved PL measurements by using an imaging spectrometer coupled to a streak camera with a time resolution of $\approx$1.8 ps (see Supporting Information and Methods). When pumping at fluences below threshold, the PL from the doped polymer layer shows a decay time of about $\tau_B=30$ ps, both on the plasmonic array (red dots in Fig. 3a) and on the bare glass substrate (green dots in Fig. 3a). Differences on the lifetime between the molecules laying on the bare substrate and those coupled to the plasmonic array would be expected as a consequence of the Purcell effect or non-radiative quenching. However, in our experiment the PL emission mainly originates from molecules spread in the whole polymer layer height (260 nm), not all coupled to the SLR, the intensity of which is mainly localised around the metallic nanostructures. As the excitation power increases above the threshold, the decay time at the energy of the PEP reduces by one order of magnitude, to about $\tau_A =$ 3 ps (blue dots in Fig. 3a). The shortening of the emission lifetime above threshold is the result of the effective scattering from the exciton reservoir to the bottom of the lower polariton branch and the short cavity photon lifetime compared to the non-radiative polariton decay rate.

We investigated further the time resolved emission above threshold by fitting the emission spectra of Fig. 3b at each time with a Gaussian peak profile (see  Supporting Information and Methods), and extracting the relative peak energy, as shown in Fig. 3c ($\approx$1 ps time steps are used). The corresponding fitted linewidths are also reported in Fig. 3d. An instantaneous blueshift as large as 2.5 meV appears when the system is excited. This blueshift is due to the mutual PEPs interactions and those of PEPs with the exciton reservoir.\cite{Daskalakis:2014ex} This is a clear signature of polariton condensation which, differently from photon lasing, manifests a density dependent energy shift \cite{}: the exciton reservoir depletes in time and the energy shows a continuous redshift towards the vacuum state (Fig. 3c) during the condensate's formation and decay. Indeed, as can be seen from the narrowing of the linewidth (Fig. 3d), the condensate is formed some ps after the excitation pulse, however the system continues to redshift due to the further reduction of the total population in the reservoir. This behaviour is fast enough to exclude any possible heating related effects which in any case would show a blue shift rather than a redshift with time and it is similar to what has been already observed in exciton-polariton condensates with inorganic semiconductor microcavities.~\cite{DeGiorgi:2014prl} This temporal dynamic not only demonstrates the presence of reservoir-PEPs interactions, but also shows that plasmon-exciton coupling still holds while condensation occurs and excludes photon lasing processes that may appear in the weak coupling limit if the coupling strength saturates.\cite{Yamamoto2010} Moreover, taking into account the absorption coefficient of the dye and the number of photons at the threshold power, we estimated the interaction constant $g = \Delta E/ N$ being $\approx 2 \cdot 10^{-23}~ eV ~ cm^{3}$ , where $\Delta E$ is the energy blueshift and $ N$ is the density of the initial excited electron-hole pairs. Considering the dilution of the dye this is in accordance with previous estimation of Frenkel excitons in organic semiconductors\cite{Daskalakis:2014ex, Lerario:2017}.

\begin{figure}
\begin{center}
\includegraphics[width=3.5in]{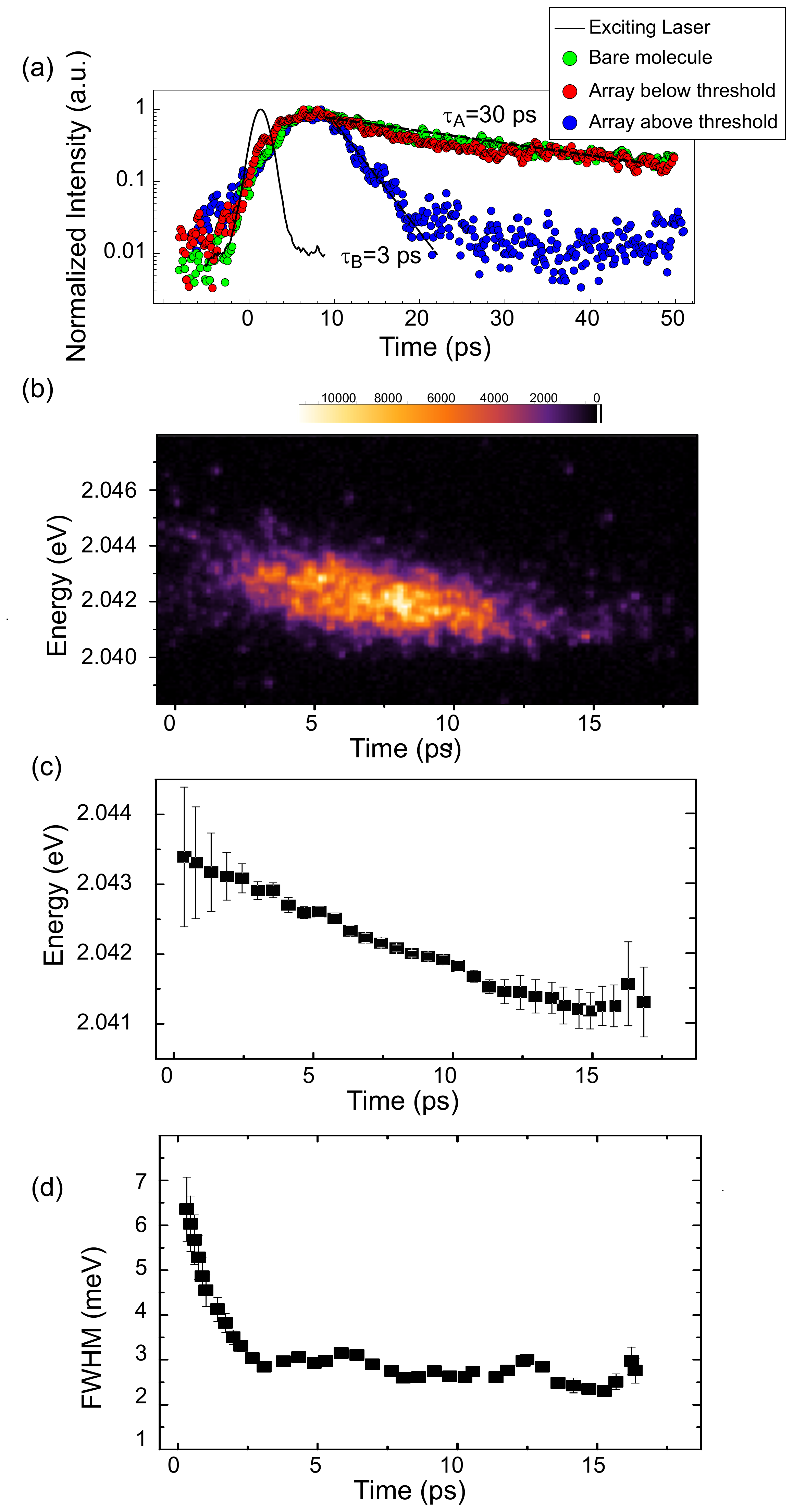}
\end{center}
\caption{\textbf{PEP temporal dynamics and interactions-driven blueshift.} (a) Time-resolved PL of the sample, below (red dots) and above (blue dots) condensation threshold. The green dots show time-resolved PL intensity for the bare polymer layer doped with molecules whereas the solid black line shows the temporal profile of the exciting laser pulse. (b) Time- and energy-resolved PL above PEP condensate threshold, as measured on the streak camera. (c) Emission peak energy at different time delay as obtained from Gaussian fit of the TRPL PEP condensate trace at different delay times. (d) Linewidth of the emission as extracted by a Gaussian fit of the time trace as a function of time.}
\label{fig:Fig3} 
\end{figure}

\section*{Spatial and temporal coherence}

One of the most important characteristics of a polariton condensate is the macroscopic spatial coherence. In PEP systems, spatial coherence of the order of few microns has been already reported in the linear regime,~\cite{BellessaPRL2009,PTormaPRL2014} and was described in terms of plasmon-exciton hybridization induced by strong coupling. On the other hand, in standard plasmonic lasing, larger coherence has been observed but always inside the excitation spot.\cite{Hoang:2017}.

In our nonlinear system, the 2D emission image of the sample above threshold retrieves all the information about spatio-temporal correlations of the PEP condensate, that can be extracted by using interferometric techniques. In particular, we have employed a Michelson interferometer, schematically shown in Fig. S2 of the Supporting Information, with the sample non-resonantly excited with a 20 $\mu m$ pulsed laser spot (E=2.48 eV, 100 fs pulse duration, 1 kHz repetition rate, as detailed in the Methods section). The corresponding emission image was collected with an objective lens and separated with a beam splitter along two perpendicular optical paths, which define the two arms of the interferometer. The two images, rotated by 180 degrees with respect to each other around the autocorrelation point (as shown in the sketch of Fig. S2 in the SI), are superimposed and interfere at the entrance slits of a monochromator equipped with a CCD camera. 

Since the sample is excited non-resonantly, the excitons relax incoherently into the PEP band, with a phase that is not imposed by the laser pump. At sufficiently high PEP densities, phononic relaxation from the exciton reservoir rapidly populates the long-living polariton state at $k_{x}=0$ $\mu m^{-1}$, leading to bosonic stimulated scattering and, finally, to condensation. By measuring the visibility of the interference fringes, we can thus obtain a complete spatial reconstruction of the first-order correlation function $g^{(1)}(r_1,r_2,\Delta t)$ of the condensate (see Supporting Information), at each temporal delay $\Delta t$ between the pulses in the two arms of the interferometer:

\begin{eqnarray}
g^{(1)}(\bf{r_1},\bf{r_2}, \Delta \normalfont{t})=\frac{<\Psi^*(\bf{r_1},\normalfont{0}) \Psi(\bf{r_2},\Delta \normalfont{t})>}{\sqrt{<\Psi^*(\bf{r_1},\normalfont{0}) \Psi(\bf{r_1},\normalfont{0})> <\Psi^*(\bf{r_2},\Delta \normalfont{t}) \Psi(\bf{r_2}, \Delta \normalfont{t})>}}\;,
\end{eqnarray}
where $\Psi^*$ and $\Psi$ are the creation and annihilation operators of the polaritons at the space-time point $(\bf{r},\normalfont{t})$.

A typical interference pattern, measured at a pumping power $P=1.2P_{th}$ and $\Delta t=0$, is shown in Fig. 4a, with the maximum fringes visibility in the center of the image (\textit{i.e.}, the autocorrelation point, $\bf{r}=\bf{r_0}$). The spatial map of the $g^{(1)}$$(\bf{-r},\bf{r},\normalfont{0})$ is displayed in Fig. 4c, as calculated from Fig. 4a, while the profile along the black dashed-line starting from the autocorrelation point is displayed in Fig. 4e. By fitting the experimental decay with an exponential function (see Supporting Information for details), a coherence length of $L_x\simeq100~\mu m$ is obtained. It is worth noting that the region of the condensate  extends to much longer distances than the pump spot size (up to four/five times as shown in Fig. S3), mediated by the SLRs spreading in the periodic array. 

In addition to the presence of 1D long-range spatial correlations along the stripes, one could wonder if, despite the disorder, the PEP condensate can still manifest the wholly 2D nature of SLRs. In order to verify this property, the $g^{(1)}$ along the direction perpendicular to the emitting stripes is measured in another position by rotating the sample, as displayed in Figs. 4d and 4f. We find that, regardless of the spatial fragmentation of the condensate, there is a high degree of coherence, which is maintained also between different stripes. In particular, by fitting all the $g^{(1)}$ maxima positions along the stripes, an exponential decay is still obtained, shown as a red curve in Fig. 4f, with a coherence length of $L_y \simeq 120~\mu m$. This value is very similar to the one obtained for the \textit{x}-direction, which demonstrates the 2D nature of the PEP condensate. In Fig. 4b, we clearly observe that by increasing the excitation power, both the intensity and the coherence length increase, as shown in Fig. 4b with black and red dots, respectively, manifesting the spontaneous buildup of a global phase at the PEP condensation energy due to the phase coherent Bose stimulated scattering process. Since the LP branch below threshold has a very low emission intensity, the pump rate dependence of the coherence length and the emission intensity have been estimated only above threshold (Fig. 4b). 

It is worth nothing that for 2D systems above a finite temperature, $T_{BKT}$, and close to equilibrium the Berezinsky-Kosterliz-Touless (BKT) transition characterized by a quasi-long-range-order, with an algebraic decay of coherence should be observed.\cite{Caputo:2017BKT} However, pumping and dissipation in our PEP system play a major role as compared to thermalisation and interactions. Due to the extremely short LSPs lifetime in plasmonic-based condensates, which make them strongly out-of-equilibrium, it is not surprising that we do not find a BKT transition, but rather an exponential loss of spatial coherence.~\cite{KenaCohen:2015}

\begin{figure}
\begin{center}
\includegraphics[width=6.5in]{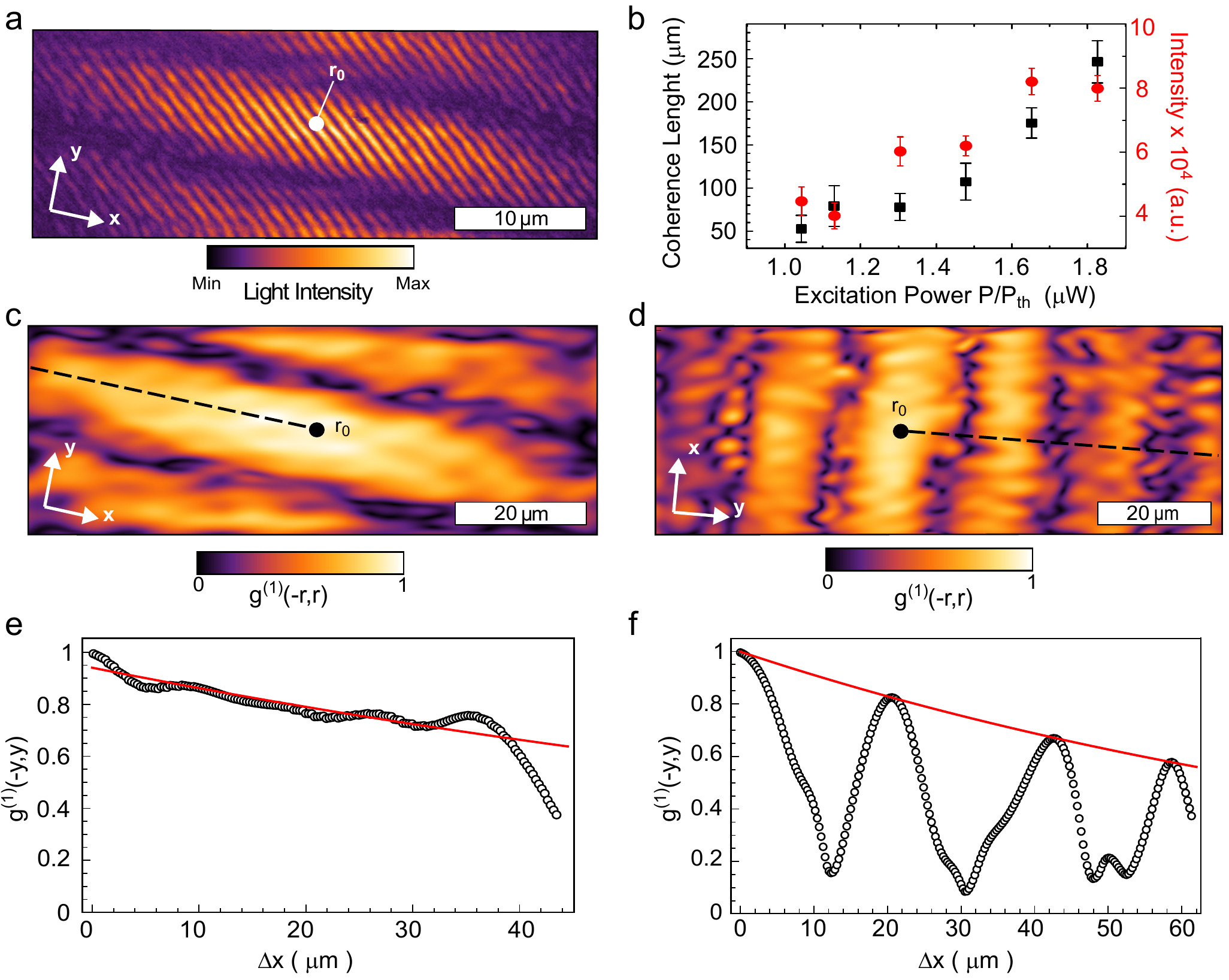}
\end{center}
\caption{\textbf{Two-dimensional spatial coherence of the PEP condensate.} (a) Experimental interferogram. (b) Emission intensity and coherence length along the x-direction as a function of the excitation power. (c) Map of first order correlation function. The dot at the centre indicates the autocorrelation point $\bf{r_0}$. (e) Profile of the $g^{(1)}(-x,x,0)$ extracted along the black dashed line in (c), parallel to the \textit{x}-axis. (d) Map and (f) profile along the dashed line, parallel to the y-axis, of the first order correlation function, in the configuration with vertical PEP condensate stripes. The red line shows the exponential fit to the coherence data showing a decay lenght of about 100 $\mu m$ for both directions.}
\label{fig:Fig4} 
\end{figure}

Finally, we also measured the coherence time of our PEP condensate as shown in Fig. 5a for three different delays on an individual emitting stripe. The $g^{(1)}$($\bf{r}=0$, $\Delta$t), at the autocorrelation point, is then calculated and plotted in Fig. 5b, displaying a loss of coherence with a decay time of $t_c \simeq 1.7$ ps, as obtained by fitting the experimental points with a quasi-Gaussian decay function (red line, see Supporting Information). We expect this coherence time to be underestimated due to the pulsed nature of the excitation. In fact, as shown in the sketch of Fig. 5b, when one of the arms of the interferometer is delayed with respect to the other, only a partial temporal superposition of the emission signals coming from the two arms can interfere, while an increasing background signal from not-overlapping pulses (shown as gray shaded regions in the sketch) reduces the overall visibility of the fringes.

\begin{figure}
\begin{center}
\includegraphics[width=3.5in]{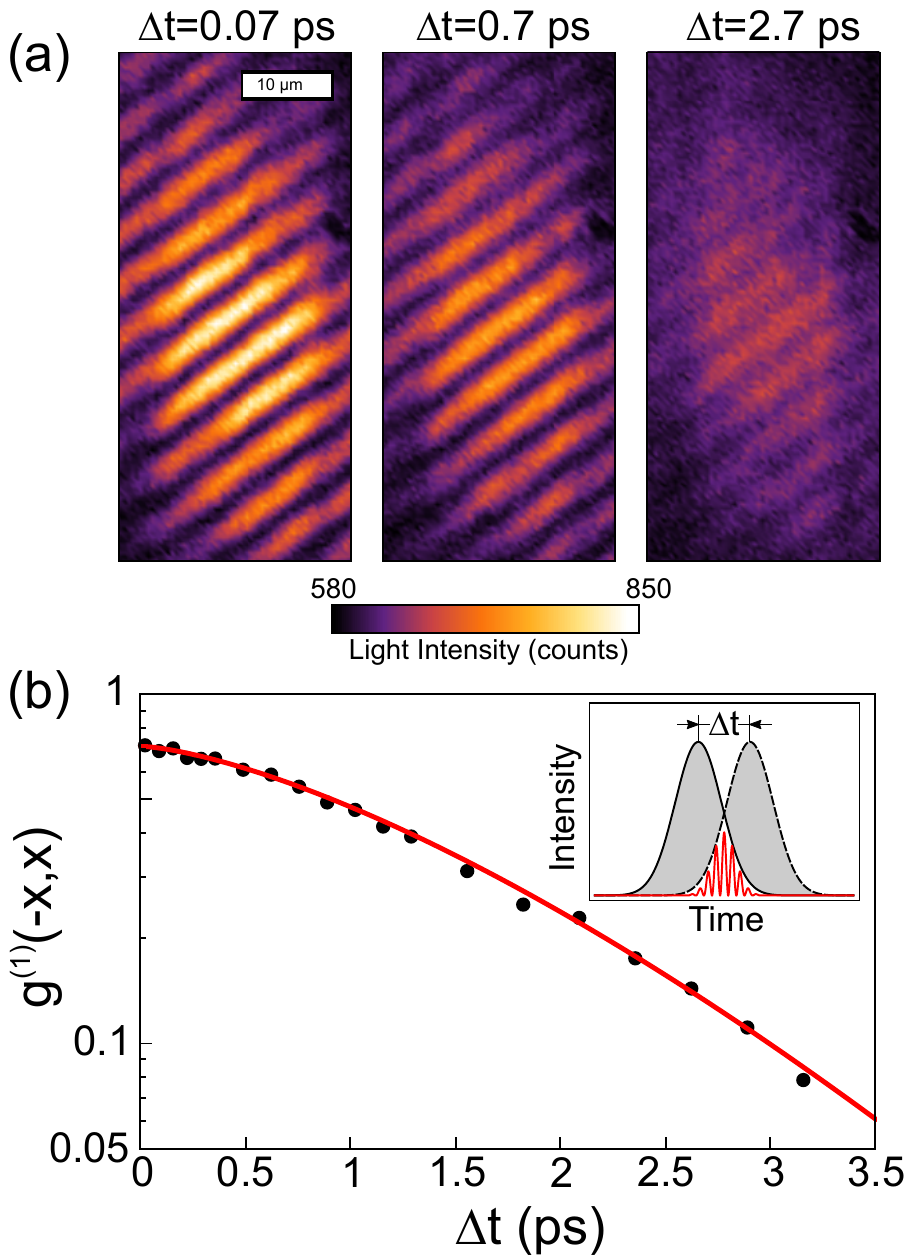}
\end{center}
\caption{\textbf{Temporal coherence of the PEP condensate.} (a) Interference pattern of an individual emitting stripe, at different delay times $\Delta t$ (0.07 ps, 0.7 ps and 2.7 ps, respectively) between the two arms of the interferometer. (b) Time decay of the coherence in the autocorrelation point, g$^{(1)}$($\bf{r_{0}}$, $ \Delta t)$. The red line shows the stretched exponential fit to the  data. Inset: sketch of the partial superposition between the signals of the two interferometer arms  resulting from the pulsed excitation and delayed in time with each other. Only the overlapped signal part  contributes to the interference.}
\label{fig:Fig5} 
\end{figure}

In conclusion we have demonstrated the formation of a non-equilibrium plasmon-exciton-polariton condensate in a lattice of metal nanoparticles supporting a long lived plasmonic mode strongly coupled to Frenkel excitons in an organic dye. Time resolved measurements reveal the ultrafast ps dynamics leading to the condensate formation, with a 2.5 meV energy blueshift of the condensate due to PEP interactions. A widely extended two-dimensional spatial coherence is also observed, showing a high degree of robustness against disorder and inhomogeneities. As a result, our system can be described as a single macroscopic state, with a coherence length longer than $100$ $\mu m$. These findings are very promising for studying properties of quantum fluids at room temperature with ultrafast dynamics, thus opening the way towards future plasmon-exciton-polariton based condensates and devices.

\section{Methods}
\subsection{Sample fabrication}
The array of silver nanoparticles was fabricated using substrate conformal imprint lithography onto a glass substrate (n=1.51). Silver nanoparticles were covered by a 8 nm thick layer of SiO$_{2}$ and a 20 nm thick layer of Si$_{3}$N$_{4}$ to protect the silver from oxidation. A 200 nm thick layer of PMMA doped with rylene dye [N,N'-Bis(2,6-diisopropylphenyl)-1,7- and -1,6-bis(2,6-diisopropylphenoxy)-perylene-3,4:9,10tetracarboximide] with 35 wt\% concentration was spin-coated onto the array.

\subsection{Optical measurements}
To measure the optical extinction and the angle-resolved PL, we have used a set of rotation stages that can rotate the sample to measure the transmission at different angles of incidence or that collects the PL at different emission angles. The transmission and PL were measured with an optical fiber and an Ocean Optics spectrometer (USB2000). For the extiction measurements we used a broadband white lamp, while for the PL measurements in Fig. 2 the sample was excited non-resonantly at normal incidence with 100 fs amplified pulses at $E_{exc}$ = 2.48 eV excitation energy with 1 kHz repetition rate. 

To study the long-range correlations on the PEP condensate, we used a Michelson interferometer. The sample was non-resonantly excited with a laser at $E_{exc}$ = 2.48 eV and 100 fs pulse width (10 kHz repetition rate, 4.5 mJ pulse energy) focused by a camera objective (3.5 cm working distance and 0.7 N.A.) into a spot of about 20 $\mu$m. In order to avoid excitation bleaching of the organic molecule, we reduced the average laser power by using a chopper with 10\% duty cycle. The PL was collected over a large area of the sample with a 40x objective and N.A. = 0.65. The real space PL maps are measured on the CCD camera by blocking one arm of the Michelson Interferometer and filtering the laser light with a long pass filter (LWP550). We used the same setup, coupled with a spectrometer equipped with a streak camera, to study the condensate temporal dynamics.

\subsection{Finite difference in time domain simulations}
The simulation of the near-field distribution shown in Figure 1 was done using a commercial package for Finite-Difference in Time-Domain simulations. A simulated volume of 380 nm $\times$ 200 nm $\times$ 2000 nm was used with periodic boundary conditions for the x- and y-direction to reproduce the periodic lattice. For the upper and lower boundaries along the z-direction, semi-infinite boundary conditions were used. The sample was illuminated with a broadband pulse incident at k = 0 $\mu m^{-1}$. We used reported values in the literature for the complex permittivites of Ag and the passivation layers of SiO$_{2}$ and Si$_{3}$N$_{4}$. \cite{palik} The permittivity of the dye doped polymer layer (shown in the S.I.) was determined by means of ellipsometry.

\begin{acknowledgement}

We are grateful of Marc A. Verschuuren for the fabrication of the samples. We also thank Femius Koenderink, Ke Guo, Dario Ballarini and Lorenzo Dominici for stimulating discussions. This research was financially supported by the Netherlands Organisation for Scientific Research (NWO) through the Industrial Partnership Program Nanophotonics for Solid State Lighting between Philips and NWO, the ERC project POLAFLOW (Grant No. 308136) and ERC-2017-PoC project ELECOPTER (grant No. 780757).

\end{acknowledgement}


%\bibliography{ACS_BIBLIOGRAPHY}
%\begin{thebibliography}

\providecommand{\latin}[1]{#1}
\makeatletter
\providecommand{\doi}
  {\begingroup\let\do\@makeother\dospecials
  \catcode`\{=1 \catcode`\}=2 \doi@aux}
\providecommand{\doi@aux}[1]{\endgroup\texttt{#1}}
\makeatother
\providecommand*\mcitethebibliography{\thebibliography}
\csname @ifundefined\endcsname{endmcitethebibliography}
  {\let\endmcitethebibliography\endthebibliography}{}

\end{document}



\section*{Molecular dye}

As an organic molecule, we used a derivative of rylene dye with the formula [N,N'-Bis(2,6-diisopropylphenyl)-1,7- and -1,6-bis (2,6-diisopropylphenoxy)-perylene-3,4:9,10-tetracarboximide] due to its photostability and the possibility of reaching high molecular concentrations within a PMMA matrix without aggregation. The absorption and emission spectra of the dye are shown in Fig. S1a. The absorption spectrum is characterized by one main peak at E = 2.24 eV and a vibronic replica at E = 2.41 eV. In Fig. S1b and c, the real and imaginary components of the refractive index of the PMMA layer doped with the rylene dye are reported, as obtained by ellipsometric measurements. These data were used to take into account the dye doped polymer dispersion in the numerical simulations.

\begin{figure}[H]
\begin{center}
  \includegraphics[width=7in]{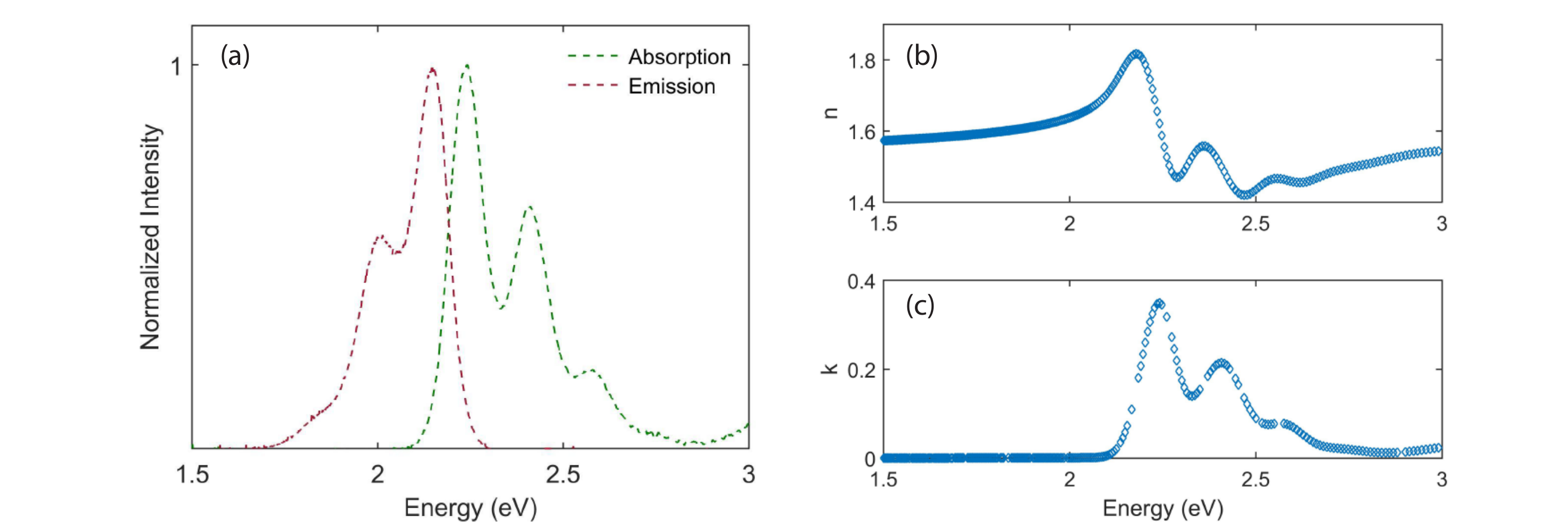} 
  \caption{Normalized absorption and emission spectra (left) and permittivity (right) of the bare rylene dye.}
\end{center}
\label{fig:FigS1} 
\end{figure}

\section*{Spatial coherence}

A sketch of the experimental setup used for the coherence measurements is depicted in Fig. S2, showing a classical Michelson interferometer scheme.

\begin{figure}[H]
\begin{center}
  \includegraphics[width=5in]{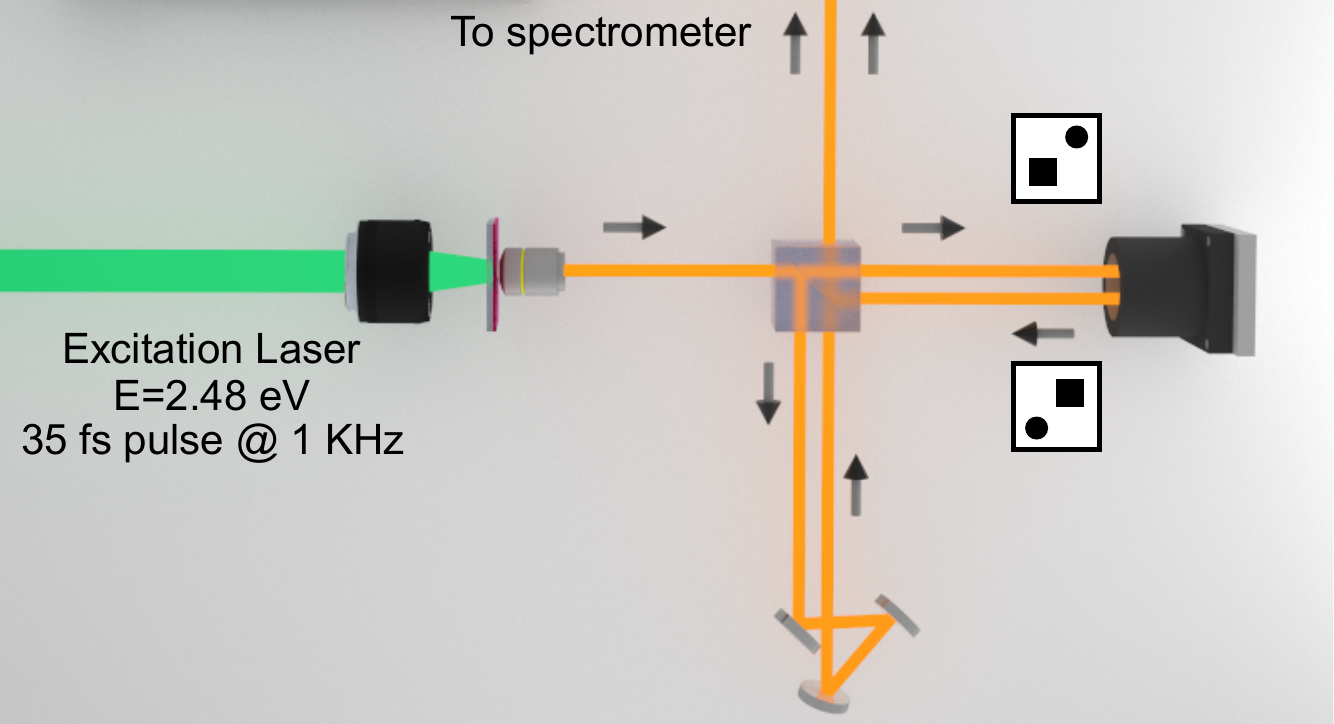} 
  \caption{Sketch of the Michelson interferometer used as experimental setup for the coherence measurements.}
\end{center}
\label{fig:FigS2} 
\end{figure}

The two-dimensional spatial map of the first order correlation function $g^{(1)}(\mathbf{r},\mathbf{-r})$ was evaluated using the fast Fourier transform (FFT) to extract the interference pattern modulation from the experimental interferogram. Selecting only the frequencies relative to the fringes modulation, allows the reconstruction of the fringes visibility. In order to normalize the measured visibility, the continuous background signal is used. The normalized visibility ($V$) and the first order correlation function have the relation

\begin{equation}
  |g^{(1)}(\mathbf{r},\mathbf{-r})|= V I_{\text{ideal}}
\end{equation}

where $I_\text{ideal}=\displaystyle (I_1+I_2)(2\sqrt{I_1I_2})^{-1}$, with $I_1$ and $I_2$ the light intensities measured on the two separated channels of the interferometer, takes into account possible small asymmetries between the two arms.

The function used for the fitting in Fig. 2d and 2f of the manuscript, is an exponential profile with the form

\begin{equation}
  |g^{(1)}(\Delta r)|= A e^{-r/b}
\end{equation}

where $A$ is a renormalization factor taking into account the experimental reduced visibility, e.g. from mechanical vibrations of the setup, and $b$ is the parameter describing the coherence length.

In Fig. S3 we report the interference patterns of the transmitted laser beam on the sample (Fig. S3a) and of the emission of the sample (Fig. S3b) filtered with a 10 nm linewidth bandpass filter centered at $\lambda = 605$ $nm$. Here a clear difference appears, in terms of the overall spatial extension of coherent emission from the sample. While the exciting laser shows a larger degree of coherence fully concentrated in the 20 $\mu m$ diameter laser spot, the coherent emission from the sample is found to be lower but more spatially extended, coming from an area much larger than the excitation spot.

\begin{figure}[H]
\begin{center}
  \includegraphics[width=5in]{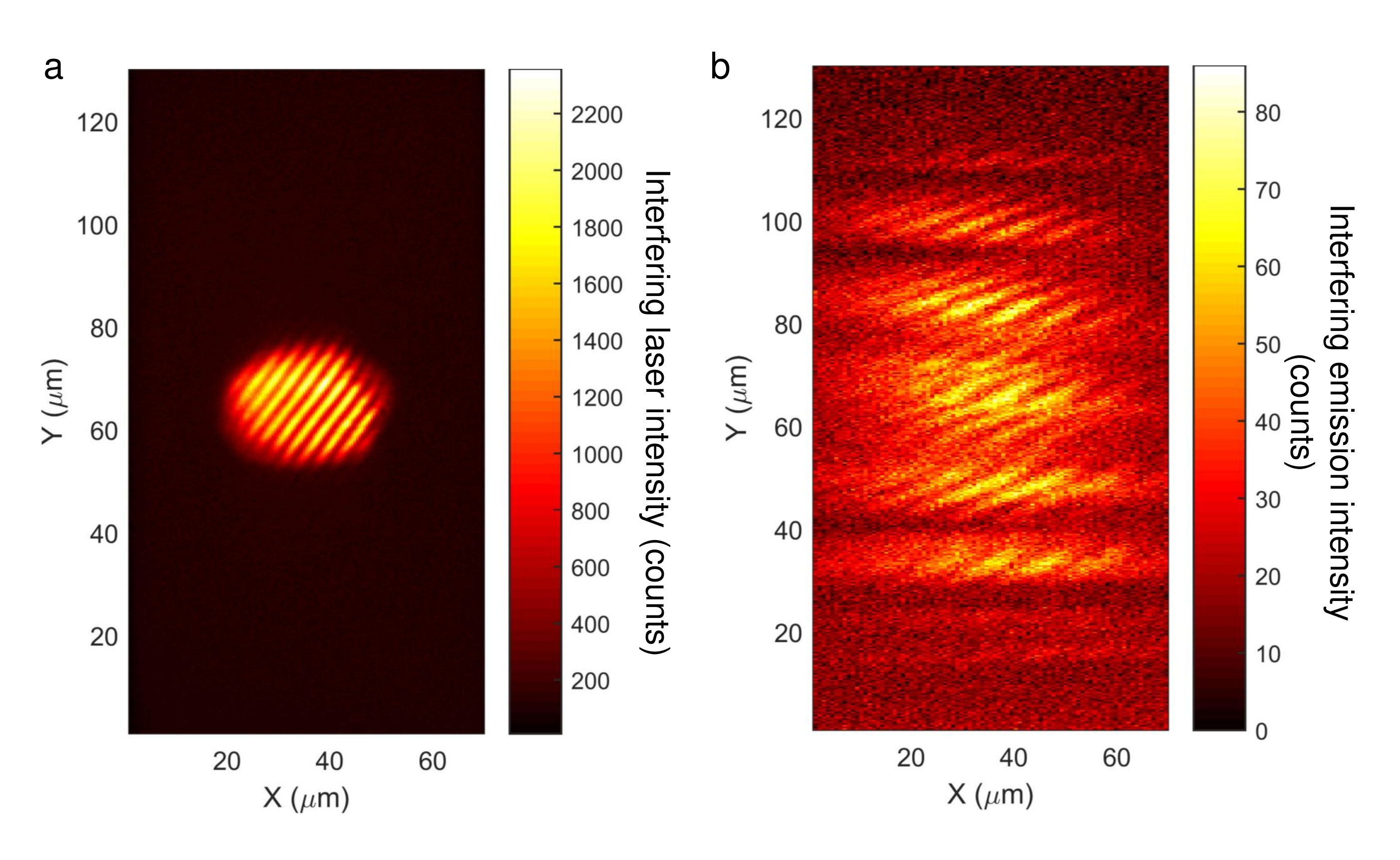} 
  \caption{Real space interference pattern from (a) the 20 $\mu m$ diameter laser spot transmitted through the sample and (b) the sample emission above threshold. The coherent emission of the sample comes from an area much larger than the excitation spot.}
\end{center}
\label{fig:FigS3} 
\end{figure}

\section*{Temporal coherence}

The temporal decay of coherence was evaluated from the coherence in the autocorrelation point ($\mathbf{r_0}$), when lengthening the optical path in one arm of the interferometer by means of a single axis translation micrometer stage. This, in turn, results in a relative temporal delay between the two arms, $\Delta t$, ranging from $0.07$ $ps$
to $\sim 3$ $ps$. Each measured frame is analysed by calculating the first order correlation function in the autocorrelation point, as discussed in the previous sections.

The extracted temporal decay is fitted by using a stretched exponential function as in:

\begin{equation}
  |g^{(1)}(\mathbf{r_0}, \Delta t)|= A e^{-(\Delta t/l_e)^{\beta}}
\end{equation}

where $A$ is a renormalization factor, while $l_e$ and $\beta$ are the parameters containing the temporal coherence decay length, 
given by:

\begin{equation}
\displaystyle
\langle l_t \rangle= \int_0^\infty dx e^{-(x/l_e)^{\beta}} = \frac{l_e}{\beta} \Gamma \left(\frac{1}{\beta} \right)
\end{equation}

where $\langle l_t \rangle$ is the temporal decay length, $l_e$ is the renormalization factor scale of the temporal axis,
$\beta$ the exponent of the stretched exponential and $\Gamma$ the gamma function. From the fit shown in Fig. 5b of the manuscript, the extracted $\beta$ is $\approx 1.5$.

\section*{Time resolved measurements}

The time resolved experiments have been performed by non-resonantly exciting the sample with a pulsed  laser of 100 fs pulse width, and collecting the emission on a streak camera. Fig. S4 shows the temporal profile of the laser pulse (in log scale) measured on the streak camera, which defines the time resolution of our set up.
By fitting the laser profile with a gaussian function (red line), a FWHM of 1.8 ps is obtained. 
The peak energy of the plasmon exciton polariton condensate at different delay times has been extracted by fitting the emission spectra at each time with a Gaussian peak profile. In Fig. S5, we show some of these fitted spectra at t = 0.5 ps (a), t = 7.2 ps (b) and t = 13.6 ps (c).

\begin{figure}[H]
\begin{center}
  \includegraphics[width=3in]{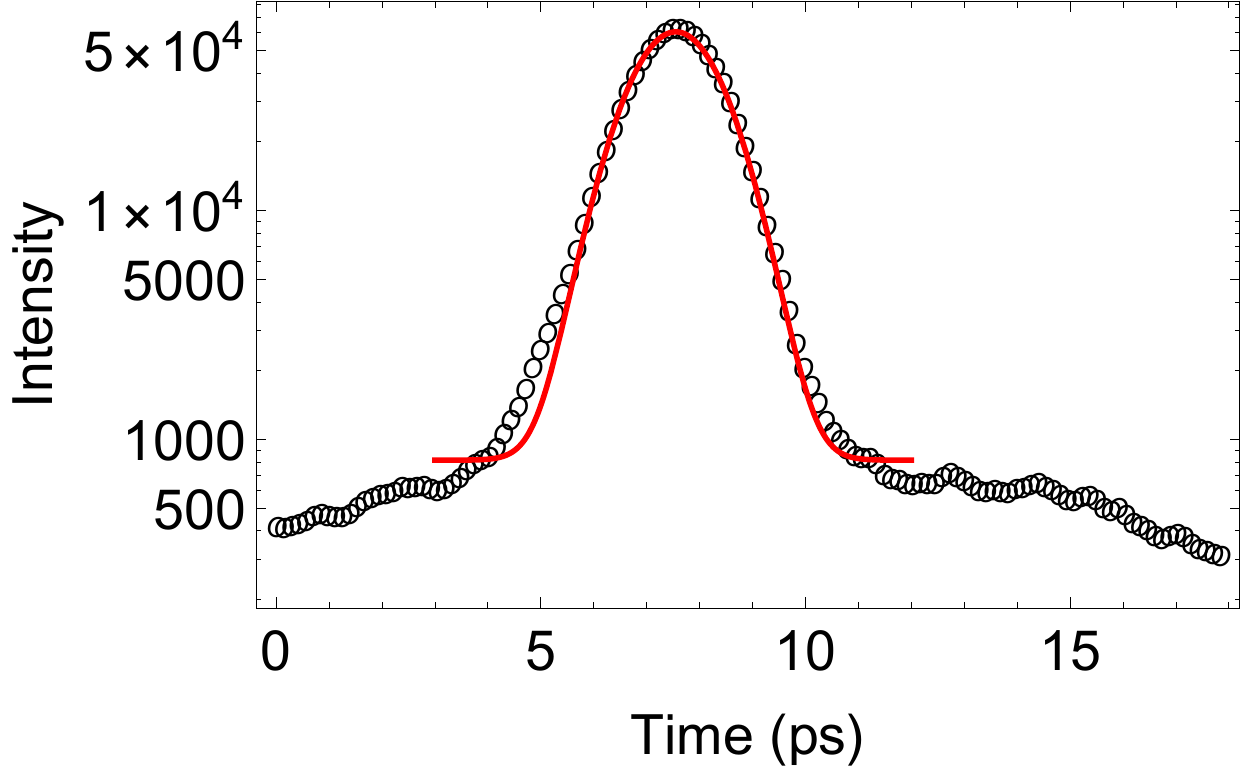} 
  \caption{Temporal profile of the 100 fs exciting laser pulse, measured on the streak camera setup. The red line shows the gaussian fit of the data with a FWHM of 1.8 ps.}
\end{center}
\label{fig:FigS4} 
\end{figure}

\begin{figure}[H]
\begin{center}
  \includegraphics[width=4in]{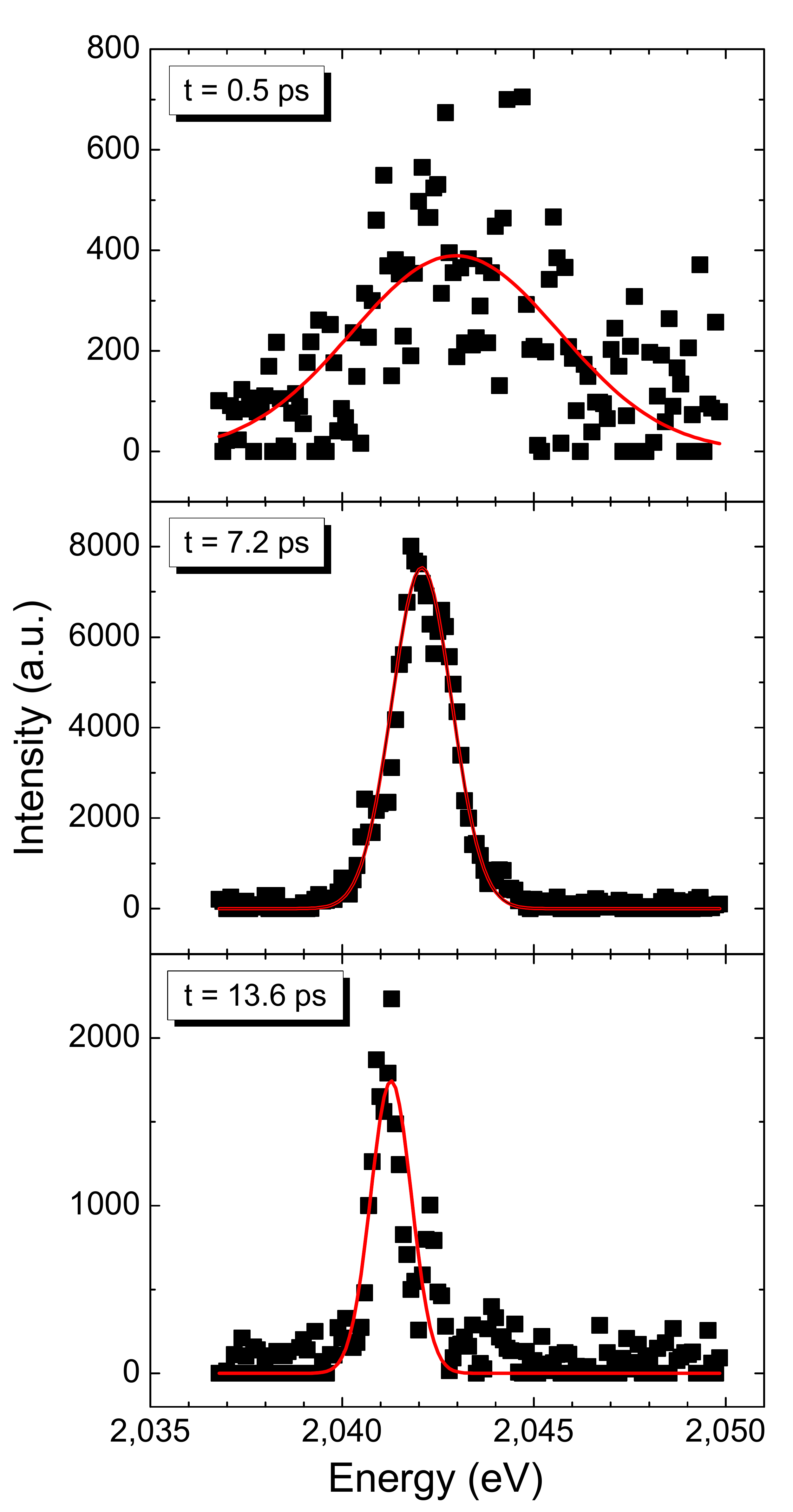} 
  \caption{Plasmon-exciton-polariton spectra at delay times of t = 0.5 ps (a), t = 7.2 ps (b) and t = 13.6 ps (c) as extracted by the time resolved emission spectrum of Fig. 3b of the main text. The red lines shows the gaussian fit of the data.}
\end{center}
\label{fig:FigS5} 
\end{figure}